\documentclass{ws-ijmpe}

\usepackage{graphicx} % do not change
\newcommand{\sqrts}{$\sqrt{s}$}
\newcommand{\pT}{$p_{T}$}

\begin{document}

\markboth{Authors' Names}{Instructions for  
Typing Manuscripts (Paper's Title)}

%%%%%%%%%%%%%%%%%%%%% Publisher's Area please ignore %%%%%%%%%%%%%%%
\catchline{}{}{}{}{}
%%%%%%%%%%%%%%%%%%%%%%%%%%%%%%%%%%%%%%%%%%%%%%%%%%%%%%%%%%%%%%%%%%%%

\title{JET AND  UNDERLYING EVENT MEASUREMENTS IN P+P COLLISIONS AT RHIC
}

\author{\footnotesize HELEN CAINES for the STAR Collaboration}

\address{Physics Department,Yale University, New Haven, CT 06520, U.S.A\\
helen.caines@yale.edu}

\maketitle

\begin{history}
\received{(received date)}
\revised{(revised date)}
%\accepted{(Day Month Year)}
%\comby{(xxxxxxxxxx)}
\end{history}

\begin{abstract}
The physics of hadron-hadron collisions is very complex involving both perturbative and non
perturbative QCD. It is  imperative to study p-p collisions in  detail  
to provide a  variety of measurements against which the theoretical calculations can be tested. 
Direct jet measurements, for instance, help address fundamental questions of the fragmentation 
process. They also form a critical baseline for comparisons of results from heavy-ion 
studies, where  fragmentation functions are expected to be modified due to interactions  with the hot and dense medium. Finally, it is important to   understand  how the beam-beam remnants, multi-parton interactions, and initial- 
and final-state radiation combine to produce the particles observed in the underlying event. I present results from p-p collisions at 200 GeV 
collisions as measured by the STAR experiment.  
\end{abstract}

\section{Introduction, the Dataset and the Analysis}

To improve our understanding of QCD and the hadronization process we must study  the properties of both jets and the underlying event in p-p collisions. %Such studies also serve as a baseline for comparison to measurements being performed in heavy-ion collisions~\cite{JetTalks}. 
The results presented here are from preliminary studies at mid-rapidity of  p-p collisions at \sqrts  = 200 GeV by the STAR collaboration.  The 
 Time Projection Chamber and Barrel Electromagnetic Calorimeter (BEMC) are used to measure the charged and neutral particle production respectively.  The data were collected using either a minimum bias trigger or a ``jet-patch" trigger, which required E$_{T}>$8 GeV in a  $\Delta \eta \times\Delta \phi$ = 1$\times$1 patch of the BEMC. The jet-patch trigger creates a neutral energy fragmentation bias for the triggering jet, hence charged particle fragmentation functions are presented only  for the di-jet partner. Jets were reconstructed using the FastJet\cite{FastJet}  package's
 k$_{T}$ and anti-k$_{T}$ recombination and the  SISCone jet algorithms. A cut of \pT\ or $(E_{T}) $ $>$0.2 GeV/c was applied to all particles considered in the event. The jet energy  resolution is of order 15-20$\%$ for jets with  \pT $>$ 10 GeV/c~\cite{CainesQM}.  

\section{Results}

\begin{figure}[htb]
		\begin{center}
			\includegraphics[width=0.6\linewidth]{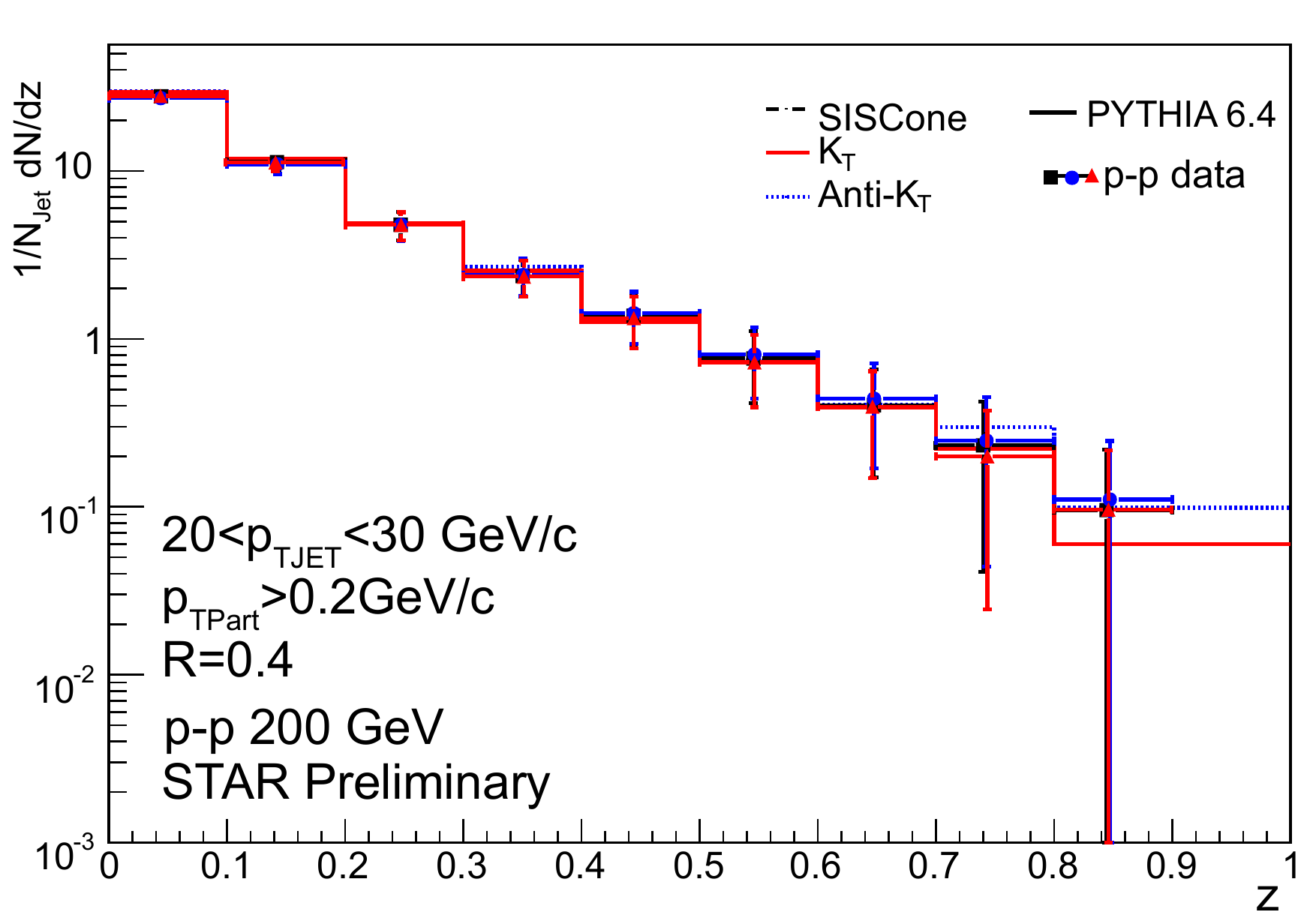}
		\end{center}
	\caption{ Color online: Charged particle, detector level FF as a function of z  for reconstructed jets  with 20$<$ \pT$<$ 30 GeV/c, compared to PYTHIA for 3 different jet algorithms. $|\eta|<$1-R, R=0.4.}
	\label{Fig:FF04}
\end{figure}

The inclusive jet spectrum\cite{ppJet} and minimum bias $\pi$ and proton spectra\cite{ppPID} have been measured at RHIC and are in good agreement with NLO pQCD calculations over several orders of magnitude. In addition preliminary fragmentation functions, FF, of charged particles have been measured.  These data are not yet corrected to the particle level. They are therefore compared to PYTHIA 6.410~\cite{PYTHIA}, tuned to the CDF 1.96 TeV data (Tune A) at the detector level.  Detector level PYTHIA events have been passed through STAR's simulation and reconstruction algorithms. Figure~\ref{Fig:FF04} shows the results for a resolution parameter R=0.4 for the three jet algorithms. The solid curves are the predictions from PYTHIA at the detector level. It can be seen that there is reasonable agreement between the data and PYTHIA. A similar agreement is observed for R=0.7~\cite{CainesQM}. This similarity, especially for the larger resolution parameter, suggests that there are only minor NLO contributions beyond those mimicked in the PYTHIA LO calculations at RHIC energies.

While the agreement between the data, NLO calculations, and PYTHIA predictions is good for $\pi$ and protons, the theoretical descriptions  of strange particle production (K, $\Lambda$ and $\Xi$) significantly under predict the yields at intermediate to high \pT\ \cite{ppStrange}. Reasonable agreement with the strange baryon data can be produced via PYTHIA if the K factor is increased from K=1 to K=3, with the effect that the $\pi$ and proton data are no longer reproduced.  To investigate strangeness production in p-p collisions further we examine strange hadron production in jets. K$^0_S$, $\Lambda$ and $\bar{\Lambda}$ are identified via their distinctive V$^{0}$ decay topologies. Good signal-to-noise ratios are obtained via topological cuts and  studying only particles with \pT $>$ 1 GeV/c. The residual background under the mass peaks is not yet corrected for. 

\begin{figure}[b]
\begin{center}
\includegraphics[width=\textwidth]{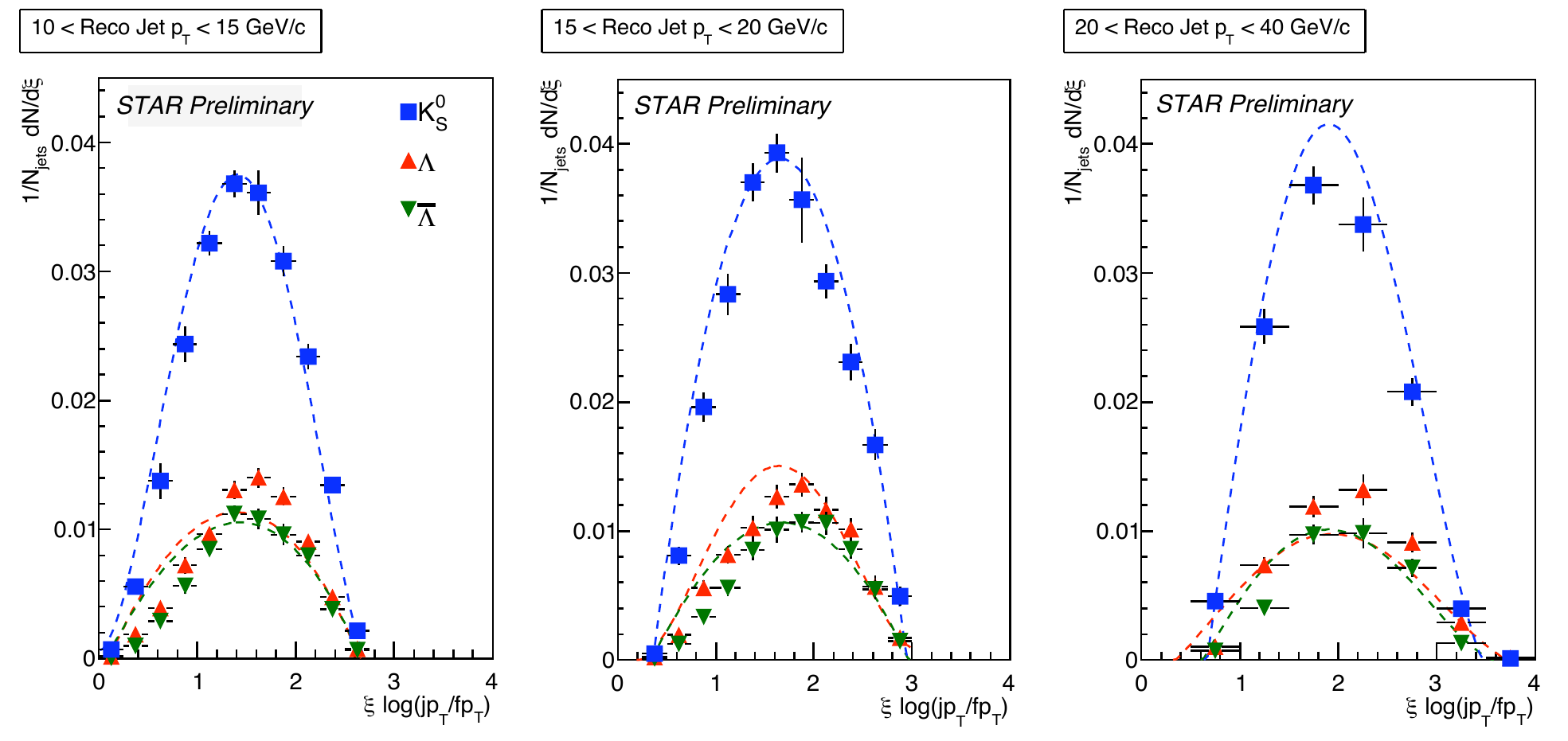}
\end{center}
\caption{Data points show strange particle FF as a function of the reconstructed jet \pT\ for R=0.4. The strange particle yields are not corrected for tracking efficiency. $jp_{T}$ and $fp_{T}$ are the reconstructed jet and fragment $p_{T}$ respectively. The curves show polynomial fits to the PYTHIA detector level predictions.} 
\label{Fig:StrangeFF} 
 \end{figure}

Jets containing these identified strange particles are studied. Figure~\ref{Fig:StrangeFF} shows the measured detector level FF; again the curves are from PYTHIA.  The uncertainties are statistical added in quadrature to the small differences obtained by comparing the three different jet finders. As with the minimum bias data, PYTHIA gives an reasonable description of the $K^{0}_{S}$ data but fails for the $\Lambda$ and $\bar{\Lambda}$. JETSET (the jet production scheme in PYTHIA) has also been shown to describe $K^{\pm}$ FF from $e^{+}+e^{-}$ collisions, for  $5 < E_{jet} < 46$ GeV \cite{FAnulli}.  Although the strange baryon  predictions from PYTHIA have the correct over-all yields in the range measured, the trend as a function of $\xi$ is incorrect, over predicting at low $\xi$ and under predicting  at intermediate $\xi$. 

The integrated $\bar{\Lambda}/\Lambda$ and $\Lambda/K^{0}_{S}$ ratios for $p_{T} > 1$ GeV/c as a function of reconstructed jet $p_T$~are consistent with the values obtained from the minimum bias inclusive spectra when the same \pT\ range is considered\cite{StrangeJets}. These PYTHIA ratios are in agreement with the data. 
%If the full \pT range is considered for the inclusive $\Lambda/K^{0}_{S}$ ratio the \pT$<$1 GeV/c range dominates and the integrated ratio of the inclusive spectra is significantly below that obtained from the jet analysis. 
This suggests that the  $\Lambda$ and $K^{0}_{S}$ spectra for $p_{T} > 1$ GeV/c have a dominant contribution from hard processes, i.e. jet production. Further studies are needed to confirm this speculation.

\begin{figure}[htb]
	\begin{minipage}{0.46\linewidth}
		\begin{center}
			\includegraphics[width=0.7\linewidth,angle=90]{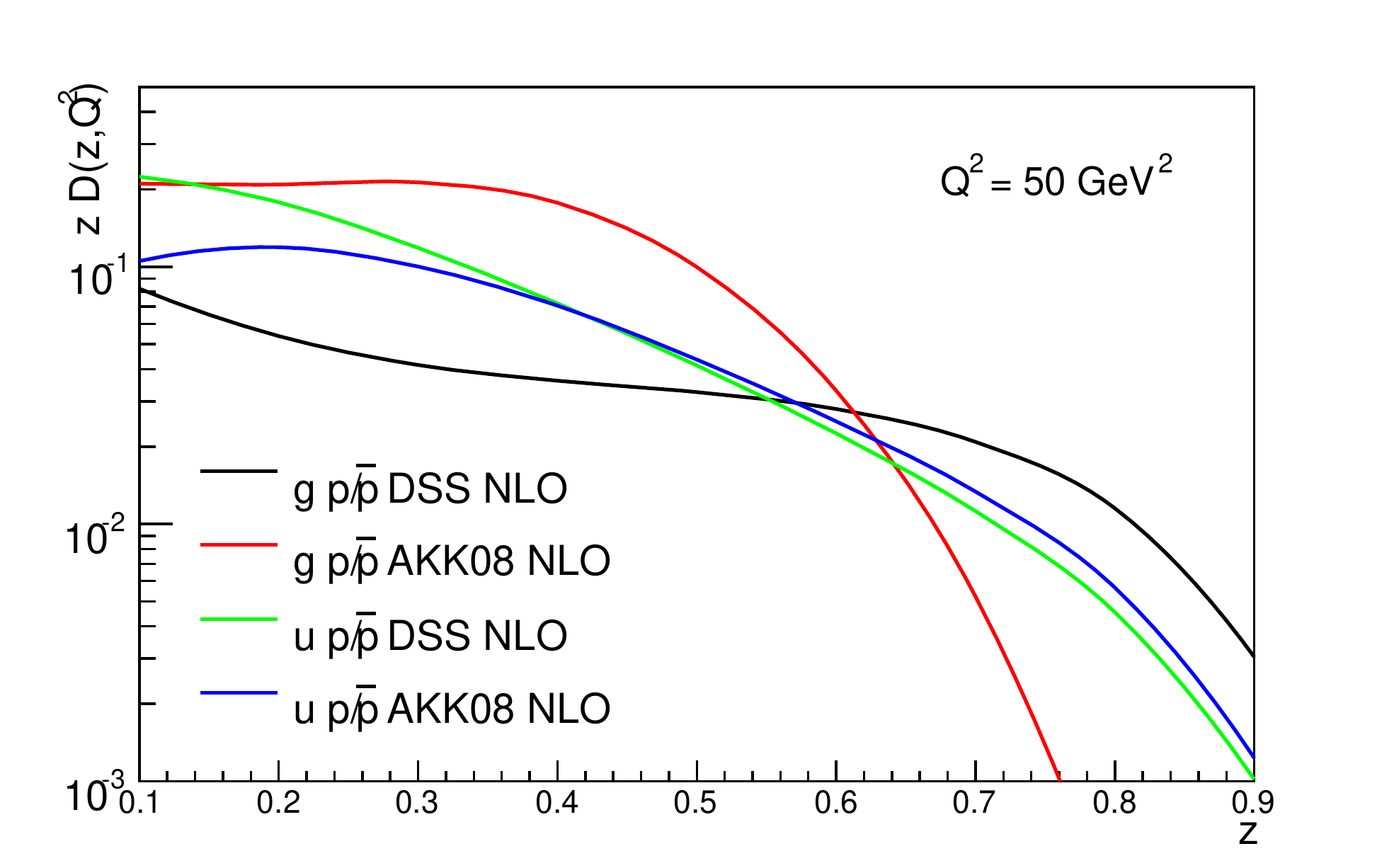}
		\end{center}
	\end{minipage}
	\begin{minipage}{0.46\linewidth}
		\begin{center}
			\includegraphics[width=\linewidth]{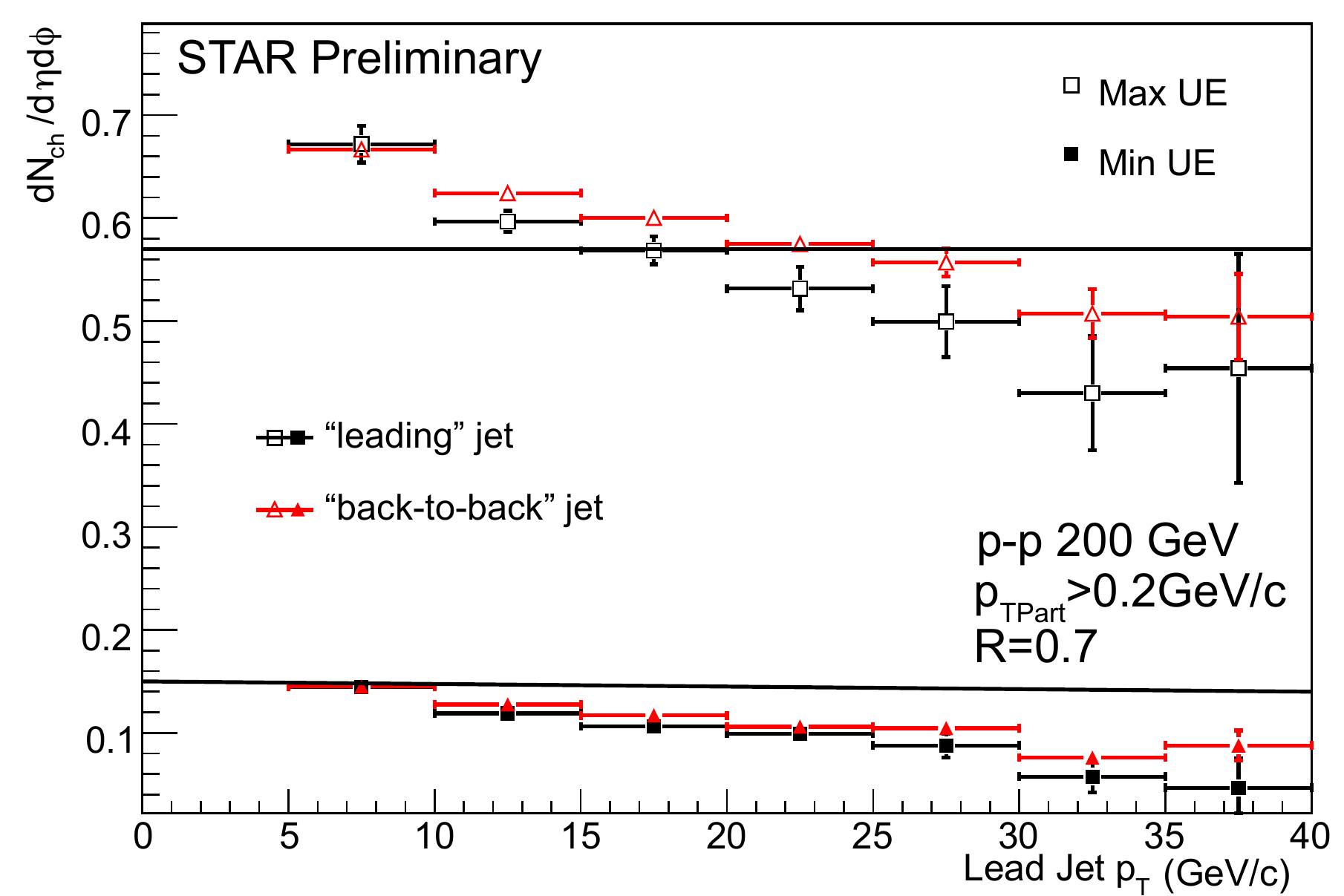}
		\end{center}
	\end{minipage}
	\caption{ Color online: Left) Calculated FF into (anti)protons for gluons and u quarks for Q$^2$ = 50 GeV$^2$. Right) The uncorrected charged particle density in the TransMin and TransMax regions as a function of reconstructed lead jet \pT, using SISCone algorithm, R=0.7}
			\label{Fig:DSSAKKUE}
\end{figure}

The detector level FF of \emph{non-leading} charged hadrons when the \emph{leading} particle is a charged hadron, $\Lambda$, or $K^{0}_{S}$ have also been examined\cite{StrangeJets}. The aim is to investigate whether gluon or quark jets can be preferentially selected by tagging on the leading particle species. Measurements by DELPHI\cite{DELPHJets} and theoretical calculations have shown that gluon jet fragmentation produces more hadrons than quark fragmentation at the same jet energy; MLLA calculations give a ratio of 9/4 for gluon/quark jet hadron multiplicities. Therefore if a leading/high \pT\ baryon is more likely to come from a gluon than a leading/high \pT\  meson, as is suggested in this paper \cite{qgJets}, we would expect the number of measured non-leading hadrons in the baryon jet to be greater than in those from meson tagged jets. However, for a given jet energy we observe that, within errors, the charged particle multiplicities are independent of the leading particle species used to tag the jet\cite{StrangeJets}. While this might seem surprising at first, given, for example, the calculations shown in \cite{qgJets}, the left plot of Fig.~\ref{Fig:DSSAKKUE} illustrates that the FF into baryons, especially from gluons, are not well constrained and thus  predictions resulting from their application have large systematic uncertainties. Both DSS\cite{DSS} and AKK08\cite{AKK08}, used to produce the left plot in Fig.~\ref{Fig:DSSAKKUE}\cite{FFWeb} give satisfactory descriptions of the available data. Further studies are therefore needed to (a) constrain the gluon FFs and (b) determine if jets can be tagged at RHIC.

In a p-p collision the underlying event (UE) is defined as everything except the hard scattering. It therefore contains contributions from beam-beam remnants, initial and final state radiation, and soft and semi-hard multiple parton interactions. Pile-up is not part of the definition of the UE. Since this study is performed at mid-rapidity, the contribution from beam-beam remnants is believed to be negligible. We follow the technique designed by CDF to study the UE\cite{CDF}.  Once the jets have been reconstructed, each event is divided into four sections defined by their azimuthal  angle with respect to the leading jet axis ($\Delta{\phi}$).  The range within $|\Delta{\phi}|$$<$60$^{0}$ is the lead jet region and that  for $|\Delta{\phi}|$$>$120$^{0}$ is the away jet area. The two  sectors with $60^{0}$$<$$|\Delta{\phi}|$$<$120$^{0}$ and $-120^{0}$$<$$\Delta{\phi}$$<$-60$^{0}$ are the transverse regions. The transverse sector containing the largest charged particle multiplicity  is  called the TransMax region, and the other is  termed the TransMin region. The TransMax region has an enhanced probability of containing contributions from the hard initial and final state radiation components.
 ``Leading" jet events, where at least one jet is found in STAR's acceptance, and  ``back-to-back"   events which are a sub-set of the ``leading" jet collection are examined. The ``back-to-back" sub-set of events  has two (and only two) found jets  with $p_{T}^{awayjet}/p_{T}^{leadjet}$$>$0.7 and $|\Delta{\phi_{jet}}|$$>$150$^{0}$. This selection suppresses hard initial and final state radiation of the scattered partons. Thus, by comparing the TransMax and TransMin regions in the ``leading" and ``back-to-back" sets we can extract information about the various components in the UE.

The right plot in Figure~\ref{Fig:DSSAKKUE} shows the measured charged particle density in the UE. First it is apparent that the UE is largely independent of the jet energy.  Second  the charged particle densities are the same within errors for the ``leading" and ``back-to-back" datasets. This again suggests that the hard scattered partons emit  very small amounts of large angle initial/final state radiation at RHIC energies. This is very different in 1.96 TeV collisions where the ``leading"/``back-to-back" density ratio is $\sim$0.65\cite{CDF}.  The black lines show the expected density assuming  events follow a Poisson distribution with an average of 0.36. The similarity of this simple simulation to the data suggests that at RHIC energies the splitting of the measured TransMax and TransMin values  is predominantly due to the statistical sampling. PYTHIA again shows satisfactory agreement with the data when Tune A is used (not shown here).

\section{Summary and Conclusions}

In summary, jet fragmentation functions have been measured in p-p collisions at \sqrts  = 200 GeV for both unidentified charged particles and strange hadrons. They will provide a stringent baseline for the measurements underway in Au-Au collisions. PYTHIA, tuned to 1.96 TeV data, shows reasonable agreement to the unidentified particle FF suggesting that the energy dependence of the underlying physics is well modeled. However the details, such as strange particle production are poorly described. Finally, studies of the UE  are underway, and show that it is largely independent of the momentum transfer of hard scattering and receives only minor contributions from initial and final state radiation from these hard scatterings.

\section*{Acknowledgements}
The author wishes to acknowledge the contributions of the Bulldog High Performance computing facility of Yale University to this work, and the support of the DoE.

\end{document}